\documentclass[preprint,showpacs,epsfig,pra]{revtex4}

\usepackage{graphicx}
\usepackage{amssymb}
\usepackage{color}

\newcommand{\figsize}{0.6}
\newcommand{\figsizetwo}{0.6}
\newcommand{\figsizethree}{0.9}

\begin{document}

\draft
\title{Whispering gallery modes in triple microdisks of triangular configurations}
\author{Jung-Wan Ryu}
\affiliation{Center for Theoretical Physics of Complex Systems, Institute for Basic Science (IBS), Daejeon 34126, Republic of Korea}
\author{Sunghwan Rim}
\affiliation{Digital Technology Research Center, Kyungpook National University, Daegu 41566, Republic of Korea}

\begin{abstract}
We study whispering gallery modes in triple microdisks of equilateral and isosceles triangular configurations. The characteristic properties of resonant modes in three microdisks on vertices of an equilateral triangle are explained by discrete rotational symmetry of the triangle. The avoided crossings of resonant modes in three microdisks on vertices of an isosceles triangle are also studied in terms of a combination of single and coupled microdisks. In addition, we propose the matrix models which well explain the resonant modes in triple microdisks.
\end{abstract}

\maketitle

\narrowtext

\section{Introduction}
Whispering gallery modes (WGMs) are optical modes that are confined inside an optical microdisk or microsphere thanks to total internal reflection \cite{McC92,Yam93,Cha96, Vah03}. These valuable modes have attracted much attention because of their high potential in many applications due to unique properties, such as ultra-high Q-factors, low mode volumes, and small sizes of resonators supporting them. When two or more microdisks supporting WGMs are located closely, WGMs can be coupled by the overlap of evanescent fields. The coupled two identical microdisks are considered as photonic molecule by the analogy with chemical molecules \cite{Bay98, Muk99, Har03, Nak05, Ish05, Bor10} and the coupled WGMs in the microdisks show various interesting behaviors such as energy-level splitting, the formation of bonding (symmetric) and antibonding (antisymmetric) states, and the oscillatory behavior of energy levels as a function of the distance between two disks \cite{Smi03, Ryu06}. Besides, coupled two non-identical microdisks exhibit avoided resonance crossing related to an exceptional point (EP) where eigenvalues and corresponding eigenstates coalesce \cite{Bor07, Ryu09}. Coupled non-identical optical devices also provide a useful platform for understanding the unique properties of non-Hermitian physics such as multiple EPs and higher-order EPs \cite{Ryu12, Hod17, Zho18}. An array of coupled microdisks called a coupled resonator optical waveguide (CROW), offer improved controlling method of the group velocity based on an evanescent-field coupling between the high-Q WGMs of individual microdisk and then applications in delaying, storing and buffering of optical signal \cite{Yar99, Poo04}. Besides multiple microdisks cavities supporting WGMs, an array of deformed microcavities can generate an enhancement in the far-field emission due to the collective operation of microcavities \cite{Kre19}.

While resonant modes in coupled two microdisks cavities have been widely studied in both cases of identical and non-identical disks, there are a few studies on resonant modes in coupled three microdisks cavities \cite{Bor06}. The resonant modes in coupled three microdisks have different properties from those in coupled two microdisks since the symmetry of the system is different from the reflection symmetry of the coupled two microdisks. If we understand the resonant modes in coupled three microdisks on a triangle, we can not only understand resonant modes in multiple microdisks on any kind of polygonal shapes due to the analogy between their symmetries but also use coupled three microdisks together with coupled two microdisks as a basic building block of complex photonic structures. In addition to these theoretical aspects, as a useful application of photonic systems, a dramatic Q-factor enhancement of coupled WGMs in triangular photonic molecules have been predicted under proper configurations \cite{Bor06} and WGM resonators have been also operated as self-referenced sensors by continuous mode-splitting, which can be explained by a perturbation method in mode coupled theory \cite{Ach19}.
In this paper, we study essential characteristics of typical resonant modes in coupled three microdisks cavities located on the triangular geometry and propose matrix models based on tight-binding model which well explains the result.

This paper is organized as follows. In Section II we study the resonant modes in a triple microdisk of an equilateral triangular configuration. Using the corresponding tight-binding models, the wavenumbers and mode patterns of resonant modes are explained in terms of symmetry. In Section III the resonant modes in a triple microdisk of an isosceles triangular configuration are considered and also explained by the corresponding tight-binding models. In Section IV we summarize our results.

\section{A triple microdisk of an equilateral triangular configuration}

\begin{figure}
\begin{center}
\includegraphics[width=\figsize\textwidth]{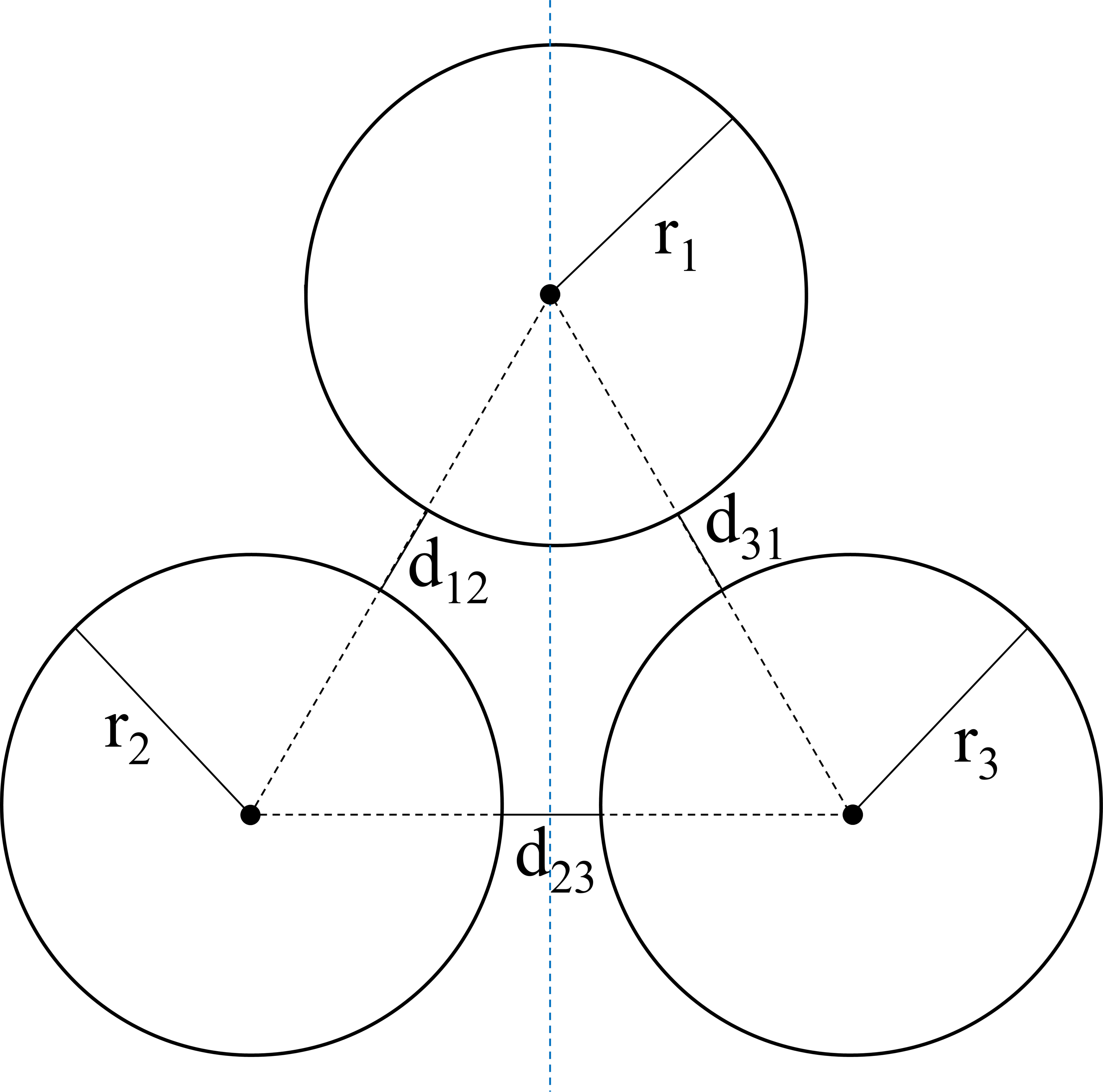}
\caption{(color online).  A triple microdisk: the centers of three microdisks are located at the three vertices of a triangle and each microdisk has radius $r_{1}$, $r_{2}$, and $r_{3}$, respectively.
The minimum distances between two microdisks are $d_{12}$, $d_{23}$, and $d_{31}$, respectively.
The blue dotted line denotes the axis of the symmetry for our numerical calculations.
}
\label{fig1}
\end{center}
\end{figure}

In this section, we study typical properties of resonant modes in three identical microdisks of an equilateral triangular configuration, i.e., the centers of each microdisk are located on the vertices of an equilateral triangle as the distances between the microdisks varies. Figure~1 shows the geometry of the system and an equilateral triangular configuration is when $r_{1}=r_{2}=r_{3}$ and $d_{12}=d_{23}=d_{31}$. 
To obtain the resonant modes numerically, we consider an infinite cylindrical dielectric cavity with a cross section of three disks of a triangular configuration. Thanks to the translational symmetry along the z-axis, one can use effective 2-dimensional dielectric cavity model, where optical modes are described by resonances or quasibound modes which are obtained by solving the Helmholtz equation,
\begin{equation}
[\nabla^{2}+n^{2}(x,y)k^{2}]\psi=0,
\label{hh}
\end{equation}
where $n(x,y)$ is the refractive index and $k$ is the wavenumber \cite{Cao15}. In this paper, the boundary element method (BEM) is used to solve the Eq.~(\ref{hh}) \cite{Wie03}. We use a dimensionless complex frequency $k=(\omega/c)R$, where $c$ is the speed of light in vacuum and $R$ is the radius of a microdisk. The real and imaginary parts of complex $k$ correspond to the frequency and decay rate of the resonant modes, respectively.
The complex-valued wave function $\psi$ represents the $z$ component of the electric (magnetic) field vector $E(B)_{z}(x,y,t) = \mathrm{Re}[\psi(x,y) \exp(-i \omega t)]$. We consider outgoing-wave condition, $\psi \sim h(\theta,k) \exp(i k r) / \sqrt{r}$, for larger $r$. This leads to modes that are exponentially decaying in time, due to the negative value of the imaginary part of $k$. The lifetime $\tau$ of the resonant modes is given by $\tau = -c \mathrm{Im}(k) / 2$ with $\mathrm{Im}(k) < 0$. The lifetime $\tau$ is related to the quality factor $Q = \mathrm{Re}(\omega) \tau$.
We focus on the transverse magnetic (TM) polarization where both the wave function and its normal derivative are continuous across the boundary and set $n(x,y)=2$ inside and $n(x,y)=1$ outside microdisks. The TM polarization means that the magnetic field is transverse to the z-axis which is the field propagation direction of the infinite cylinder. In the case of transverse electric (TE) polarization, the wave function and its normal derivative divided by $n^2$ are continuous across the boundary. It is noted that there is no significant difference between TM and TE polarizations, except for the resonant modes related to the Brewster angle in the TE polarization, which are not high-Q modes. We calculate the resonant modes using the reflection symmetry of the system about the vertical axis (blue dotted line in Fig.~1), i.e., even and odd parities.

\subsection{Resonant modes}

The resonant modes of a single microdisk are well classified by angular momentum mode index $m$ and the radial mode index $l$ \cite{Cha96}. They are twofold degenerated because of the rotational symmetry except for $m=0$ case.
In a double identical microdisk, the whispering gallery type resonant modes are fourfold degenerated when the microdisks are separated infinitely far apart. These degenerate modes are split as the distance between the microdisks decreases \cite{Ryu06}.
Likewise in a triple identical microdisk, there are corresponding sixfold degeneracy in resonant modes if the distances between microdisks are infinite. The sixfold degeneracy is lifted when the distance between the microdisks becomes finite. Therefore, the separation of these sixfold degenerate resonant modes is the essential mechanism behind the characteristic properties of resonant modes as the distance between the cavities decreases for the triple microdisk. So, it is indispensable to understand the mechanism of the separations of the sixfold degenerate modes to understand the whole picture of resonant mode behavior.

\begin{figure}
\begin{center}
\includegraphics[width=\figsizetwo\textwidth]{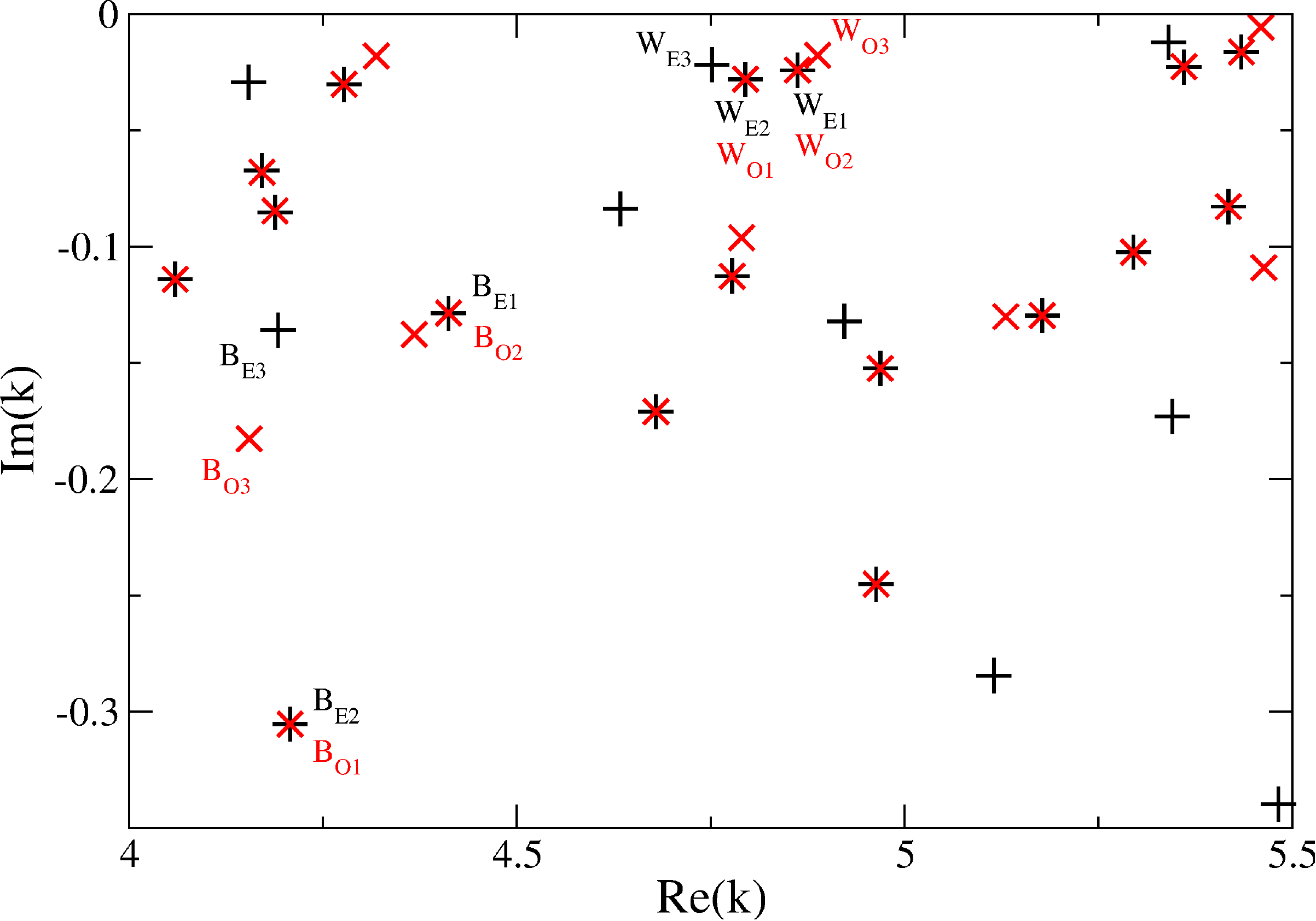}
\caption{(color online). Resonant modes in a triple microdisk of an equilateral triangular configuration. Black plus and red cross represent the resonant modes with even and odd parities, respectively.
}
\label{fig2}
\end{center}
\end{figure}

Figure~2 shows resonant modes in three identical microdisks located on the vertices of an equilateral triangle of which lengths of edges equal to $2r+d$, where $r=r1=r2=r3=1.0$ and $d=d_{12}=d_{23}=d_{31}=0.2$. Resonant modes with even and odd parities are obtained independently using BEM. The interesting point is that there are pairs of degenerated modes with different parities.
In general, for each pair of mode numbers $(\pm m, l)$ of the single cavity except for $m=0$, there are six resonant modes which are originated from the sixfold degenerate modes of infinite $d$ case. These six resonant modes can be combinations of WGMs for the case of $m > \mathrm{Re}(k)$. Figure~3 shows the mode patterns which are combinations of WGMs and we call W-modes. Figure~4 shows the mode patterns which are combinations of bouncing ball type modes which are similar to modes in a Fabry-P{\'e}rot resonator \cite{Bar13} and we call B-modes. While it is clear that the W-modes originate from WGMs in a single microdisk because of weak coupling, it is not so clear which are corresponding modes in a single microdisk of low-Q modes including B-modes because of strong coupling. If we consider the evolution of complex $k$ as $d$ varies, it is very clear that W-modes originate from WGMs in a single microdisk (not shown here).

\begin{figure}
\begin{center}
\includegraphics[width=\figsizetwo\textwidth]{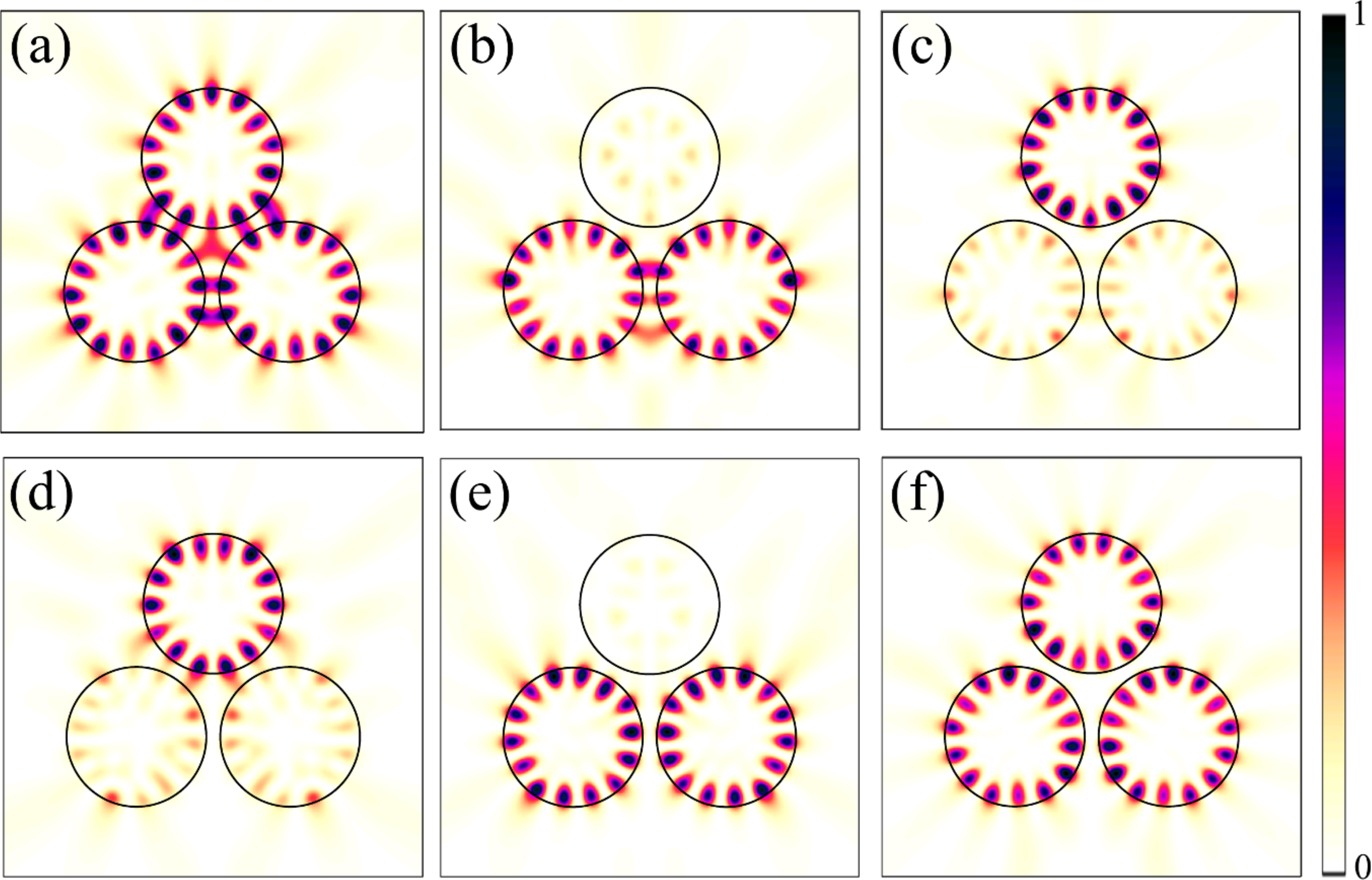}
\caption{(color online). The normalized intensity patterns of six resonant modes, (a) $W_{E3}$, (b) $W_{E2}$, (c) $W_{E1}$, (d) $W_{O1}$, (e) $W_{O2}$, and (f) $W_{O3}$, are plotted as an example of typical weakly coupled case. The first and second subscripts represent the parity and the number of microdisks where high-intensity patterns appear, respectively. 
}
\label{fig3}
\end{center}
\end{figure}

\begin{figure}
\begin{center}
\includegraphics[width=\figsizetwo\textwidth]{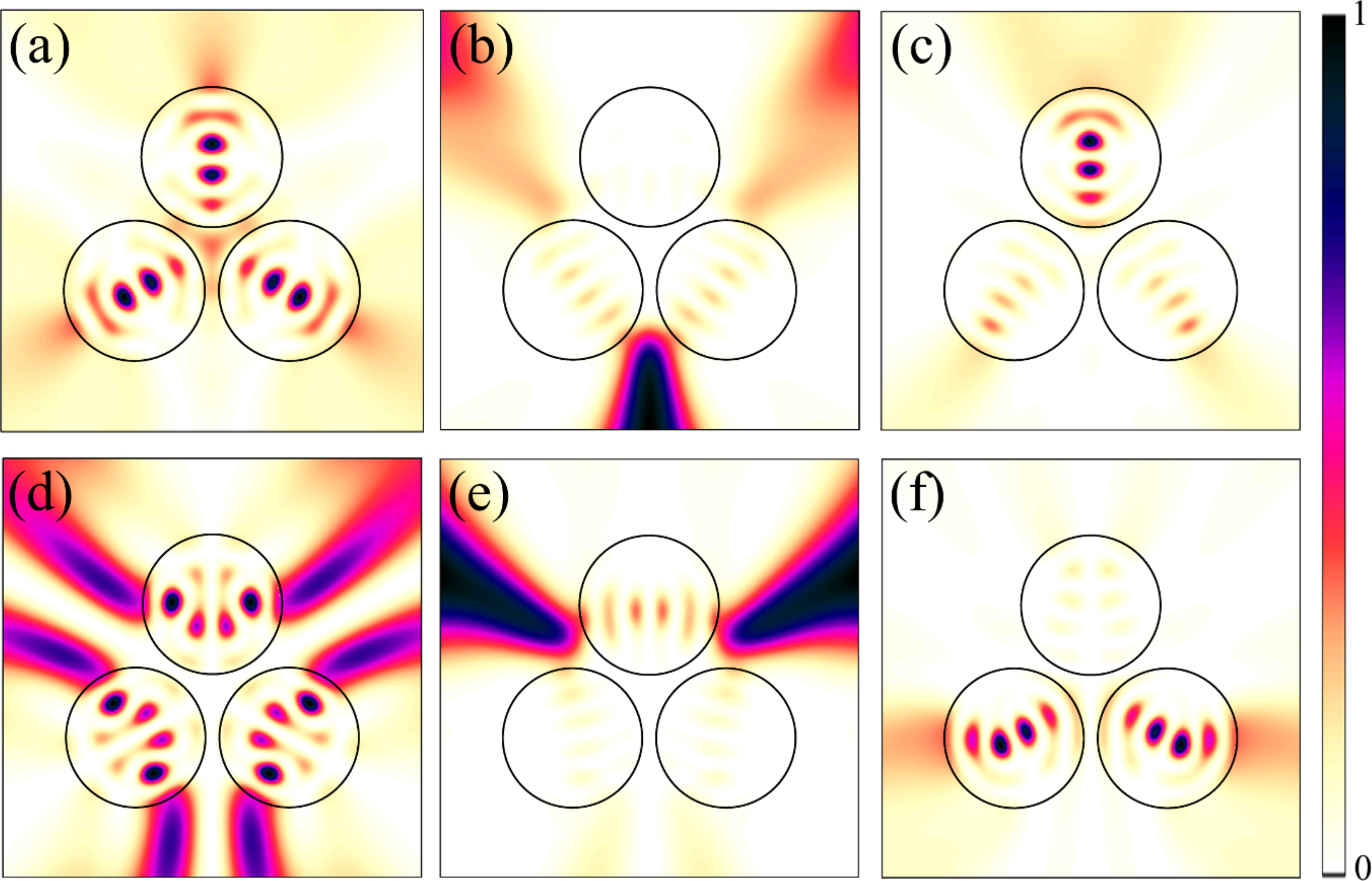}
\caption{(color online). The normalized intensity patterns of six resonant modes, (a) $B_{E3}$, (b) $B_{E2}$, (c) $B_{E1}$, (d) $B_{O3}$, (e) $B_{O1}$, and (f) $B_{O2}$, are plotted as an example of typical strongly coupled case. The first and second subscripts represent the parity and the number of microdisks where high-intensity patterns appear, respectively.
}
\label{fig4}
\end{center}
\end{figure}

There are two kinds of resonant modes in a triple microcavity of an equilateral triangular configuration. (i) The non-degenerated resonant modes: these modes show three identical patterns on each of three microdisks as shown in Fig.~3 (a) and (f) [or Fig.~4 (a) and (d)]. These mode patterns have $2\pi / 3$ rotational symmetry of the equilateral triangle. (ii) The degenerated resonant modes: these modes show high-intensity patterns on one or two microdisks as shown in Fig.~3 (b), (c), (d), and (e) [or Fig.~4 (b), (c), (e), and (f)]. These modes are symmetric about a vertical axis so we can classify them into even and odd parities. Considering the mode patterns with even parity, two rotation patterns are obtained by clockwise and counterclockwise $2\pi/3$-rotations of the even parity mode with $k$. Subtracting the two rotation patterns gives an odd parity mode with the same $k$. As a result, the even parity mode which has a high-intensity pattern in one (two) microdisk is degenerated with the odd parity mode which has a high-intensity pattern in two (one) microdisks. Superposed modes of a pair of degenerated even and odd parity modes show three identical patterns on each of the three microdisks. It should be noticed that the localization of mode patterns in a triple microdisk of an equilateral triangular configuration depends on the threefold rotational symmetry, $C_{3v}$ symmetry \cite{group}, and are independent of specific mode properties such as Q-factors. Considering the symmetry, we can use similar analysis introduced in this paper for the $C_{n v}$ symmetry with $n>3$.

\subsection{Matrix model}

To understand the characteristic properties of resonant modes and the corresponding mode patterns in the triple microdisk, we introduce a simple $3 \times 3$ matrix $H_e$ of the form
\begin{equation}
H_e = \left(
\begin{array}{ccc}
\epsilon_1 & -c_{12} & -c_{31} \\
-c_{12} & \epsilon_2 & -c_{23} \\
-c_{31} & -c_{23} & \epsilon_3 \\
\end{array}
\right),
\label{matrix_1}
\end{equation}
where $\epsilon_{n}$ corresponds to the frequency of uncoupled resonant modes on the $n$-th microdisk and $c_{lm}$ does the coupling strength between modes on the $l$-th and $m$-th microdisks, where $l,m=1,2,3$. We only consider high-Q resonant modes localized inside cavities and then basis states of $H_e$ correspond to the amplitudes of resonant modes localized in one among three microdisks, respectively, which are WGMs in each microdisk. From the symmetry of the system, we can assume $\epsilon_{1}=\epsilon_{2}=\epsilon_{3}=\epsilon$ and $c_{12}=c_{23}=c$ without loss of generality. First, we consider the case of $c_{31}=c$. The eigenvalues of $H_e$ are $\epsilon-2c$ and $\epsilon+c$. The latter is degenerated. The eigenstates are $(1,1,1)$, $(1,-1/2,-1/2)$, and $(0,1,-1)$. If we consider the geometry of an equilateral triangle as shown in Fig.~1, these eigenstates are regarded as the even, even, and odd parity states, respectively, since the parity is decided by the signs of second and third elements of eigenstates when their absolute values are the same. For example, there are high-intensity patterns in second and third microcavities in Fig.~\ref{fig3} (b) and (e) but their amplitudes have the same (even parity) and different (odd parity) signs, respectively. Next, if we consider the case of $c_{31}=-c$, the eigenvalues of $H_e$ are $\epsilon+2c$ and $\epsilon-c$. The latter is also degenerated. The eigenstates are $(1,-1,1)$, $(1,1/2,-1/2)$, and $(0,1,1)$ and then they are regarded as odd, odd, and even parity states, respectively.

Table~1 shows the correspondence between eigenvalues and eigenstates of $H_e$ and corresponding modes in a triple microdisk. For the high-Q resonant modes such as the six resonant modes of Fig.~3, the couplings between resonant modes on the microdisks are caused by tunneling, i.e., evanescent field coupling, because the modes are confined by total internal reflection. In this weak coupling case, despite the $3 \times 3$ matrix $H_e$ is simple, the matrix model well describes the high-Q resonant modes in a triple microdisk.

\begin{table}[hbt!]
\begin{center}
\begin{tabular}{|r|c|c|c|c|}
\hline
Eigenvalue & $\epsilon-2c$ & $\epsilon-c$ (deg.) & $\epsilon+c$ (deg.) & $\epsilon+2c$ \\
\hline
Eigenstate (Even) &
$\left(\begin{array}{c}
1\\
1\\
1\\
\end{array}
\right)$ &
$\left(\begin{array}{c}
0\\
1\\
1\\
\end{array}
\right)$ &
$\left(\begin{array}{c}
1\\
-1/2\\
-1/2\\
\end{array}
\right)$ &
\cr
\hline
Eigenstate (odd) &
& $\left(\begin{array}{c}
1\\
1/2\\
-1/2\\
\end{array}
\right)$ &
$\left(\begin{array}{c}
0\\
1\\
-1\\
\end{array}
\right)$ &
$\left(\begin{array}{c}
1\\
-1\\
1\\
\end{array}
\right)$ \cr
\hline
Corresponding mode & $W_{E3}$ & $W_{E2}$ & $W_{E1}$ & \cr
  & & $W_{O1}$ & $W_{O2}$ & $W_{O3}$ \cr
\hline
\end{tabular}
\caption{Eigenvalues and eigenstates of the $3 \times 3$ matrix of Eq.~(\ref{matrix_1}). The final row shows the corresponding high-Q resonant modes in the triple identical microdisk.}
\end{center}
\end{table}

It is noted that a perturbative method in mode coupled theory to explain their new WGM-based sensing technique have been developed \cite{Ach19}. According to the theory, we can see that the sixfold degeneracy in WGMs is lifted monotonically as the distances between microdisks decreases. This theory well explains not only the order of frequencies, i.e., $\mathrm{Re}(k_{W_{E3}}) < \mathrm{Re}(k_{W_{E2}}), \mathrm{Re}(k_{W_{O1}}) < \mathrm{Re}(k_{W_{E1}}), \mathrm{Re}(k_{W_{O2}}) < \mathrm{Re}(k_{W_{O3}})$, of coupled WGMs in triple microdisks (eigenvalues of $3 \times 3$ Matrix model of Eq.~(\ref{matrix_1})) but also that in double microdisks \cite{Ryu06}.

\section{A triple microdisk of an isosceles triangular configuration}

In this section, we study resonant modes in a triple microdisk of an isosceles triangular configuration in terms of avoided crossing. The radius $r_1$ of the top microdisk can be varied and $r = r_2 = r_3 = 1.0$ and $d=d_{12}=d_{23}=d_{31}=0.2$ are chosen. Except for the case of $r_1=1.0$, the $2 \pi / 3$ rotational symmetry of an equilateral triangle is broken and only one reflection symmetry about vertical axis remains. However, we note that the parity we introduced in the previous section is still valid. Based on this fact, it is convenient to consider the triple microdisk setup as a coupled system of a single size-variable microdisk and coupled two size-fixed microdisks. 

\subsection{Resonant modes}

First, assume that there is no coupling between single and coupled microdisks, i.e., the single and coupled microdisks are considered independently, the resonant modes as $r_1$ varies are shown in Fig.~5(a). We only consider WGMs of which couplings are weak. The real part of complex $k$ of WGM in a single microdisk is inversely proportional to the $r_1$ and they are degenerated with different symmetry class about a vertical axis. The real parts of complex $k$ of coupled WGMs in coupled microdisks of which radii are $r_2$ and $r_3$ are constant as a function of $r_1$ and split into four $k$ with EE, EO, OE, and OO symmetries. The former letter E (O) is even (odd) if the wave function is even (odd) with respect to the vertical axis and the latter refers to the horizontal axis which runs through the centers of coupled disks. The resonant modes with EE and EO symmetries are nearly degenerated and those with OE and OO symmetries are also nearly degenerated. The degree of splitting increases as the $d_{23}$ decreases.

\begin{figure}
\begin{center}
\includegraphics[width=\figsize\textwidth]{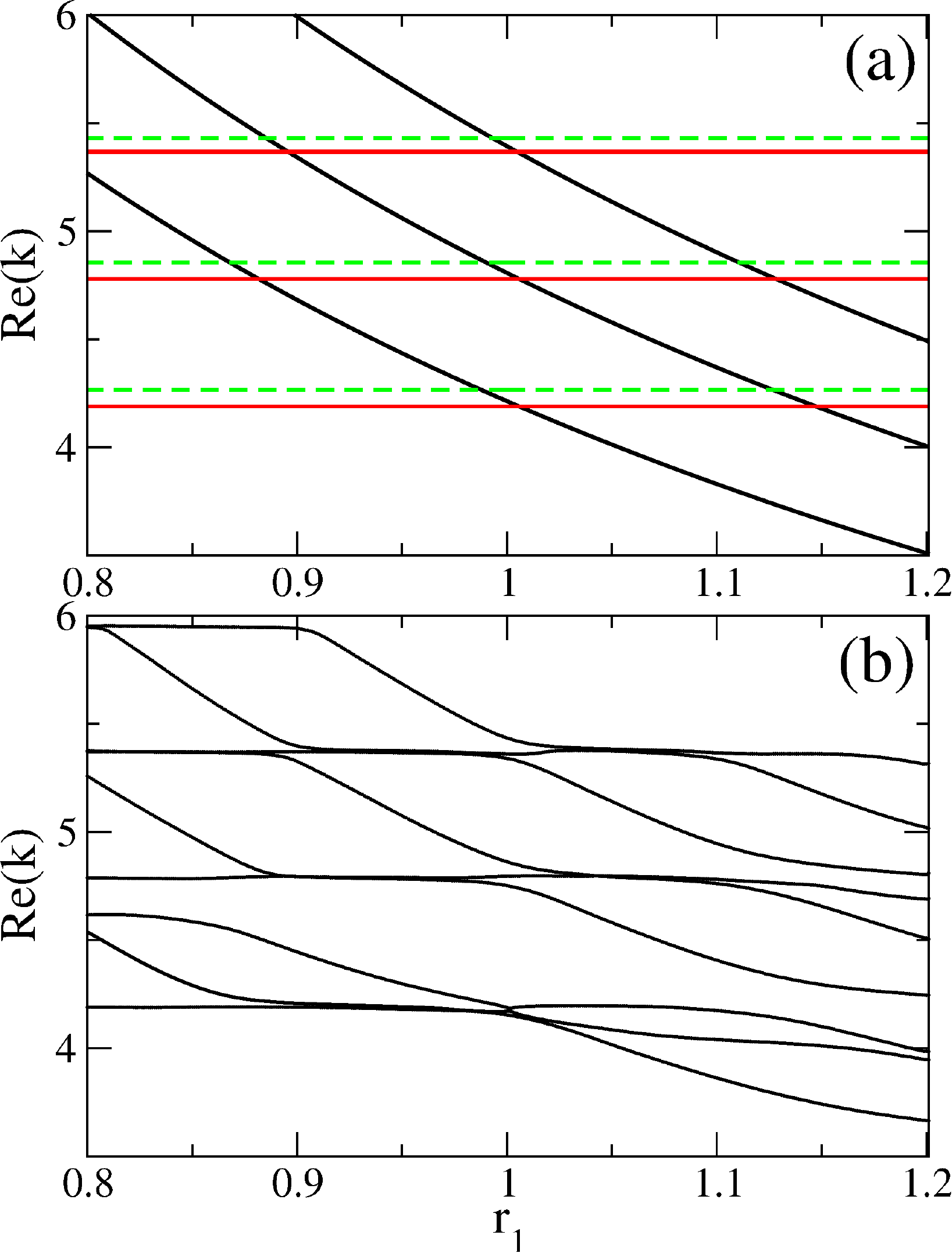}
\caption{(color online). (a) Real parts of complex $k$ as a function of $r_{1}$ on the assumption that there is no coupling between WGMs of single and coupled microdisks, i.e. $d_{12}=d_{31}=\infty$. The black (curved) lines represent degenerated resonant modes on a single microdisk. The red (straight) and green (dashed) lines represent nearly degenerated resonant modes with even and odd symmetries about a vertical axis on coupled microdisks, respectively. (b) Real parts of $k$ of the resonant modes with even parities about a vertical axis as a function of $r_{1}$ for $d=0.2$.
}
\label{fig5}
\end{center}
\end{figure}

Figure~5(b) shows the real parts of $k$ of the resonant mode with even parity about a vertical axis as a function of $r_1$ when the coupling is weak, i.e., $d$ is large enough. Since the coupling is weak, the resonant modes are similar to those of the case of resonant modes (black curved lines in Fig.~5(a)) on a single microdisk and two nearly degenerated resonant modes (red straight lines) with even parity about the vertical axis on coupled microdisks of Fig.~5(a) except for the narrow regions of real value avoided crossing (RAC) or real value crossing (RC). As the coupling gets stronger, i.e., $d$ becomes smaller, the region and gap size of RAC are wider and larger, respectively, and then the classification between the resonant modes of single and coupled microdisks becomes unclear. The resonant modes with odd parity are also similar to those of the case of resonant modes (black curved lines) on a single microdisk and two nearly degenerated resonant modes (green dashed lines) with odd parity about a vertical axis on coupled microdisks of Fig.~5(a).

Many similar shaped RACs appear periodically as shown in Fig.~5(b), where the negative slope of large real part changes into the horizontal slope, the horizontal slope of the middle real part does not change, and the horizontal slope of the small real part changes into the negative slope. The corresponding eigenstates of eigenvalues of which real parts have negative slops as functions of $r_1$ show high-intensity patterns on a single microdisk. The corresponding eigenstates of eigenvalues with constant real parts show high-intensity patterns on coupled microdisks. The RACs occur with corresponding imaginary value crossings as shown in Fig.~6 (a) and (b). The appearance of additional RACs or RCs around $r_{1}=0.85$ and $r_{1}=0.95$ occur frequently because two resonant modes are always very close as shown in Fig.~5.

Among the resonant modes with the largest real parts (green line in Fig.~6 (a)), the intensity pattern of mode-A has a single WGM on the top microdisk but the intensity pattern of mode-B has three WGMs on the all microdisks under the influence of RAC. As $r_1$ increases, the intensity pattern of mode-C has coupled WGMs. Among the middle and smallest resonant modes (red and black lines), the intensity patterns of mode-D and mode-G show coupled WGMs with the almost even and odd symmetries about the horizontal line which passes through the centers of left and right microdisks. These approximate symmetries are caused by the absence of intensity on the top microdisk. Finally, one single and two coupled WGMs at $r_1 = 0.875$ change into two coupled and one single WGMs at $r_1 = 0.925$ via mixed WGMs at $r_1 = 0.9$.

\begin{figure*}
\begin{center}
\includegraphics[width=\figsizethree\textwidth]{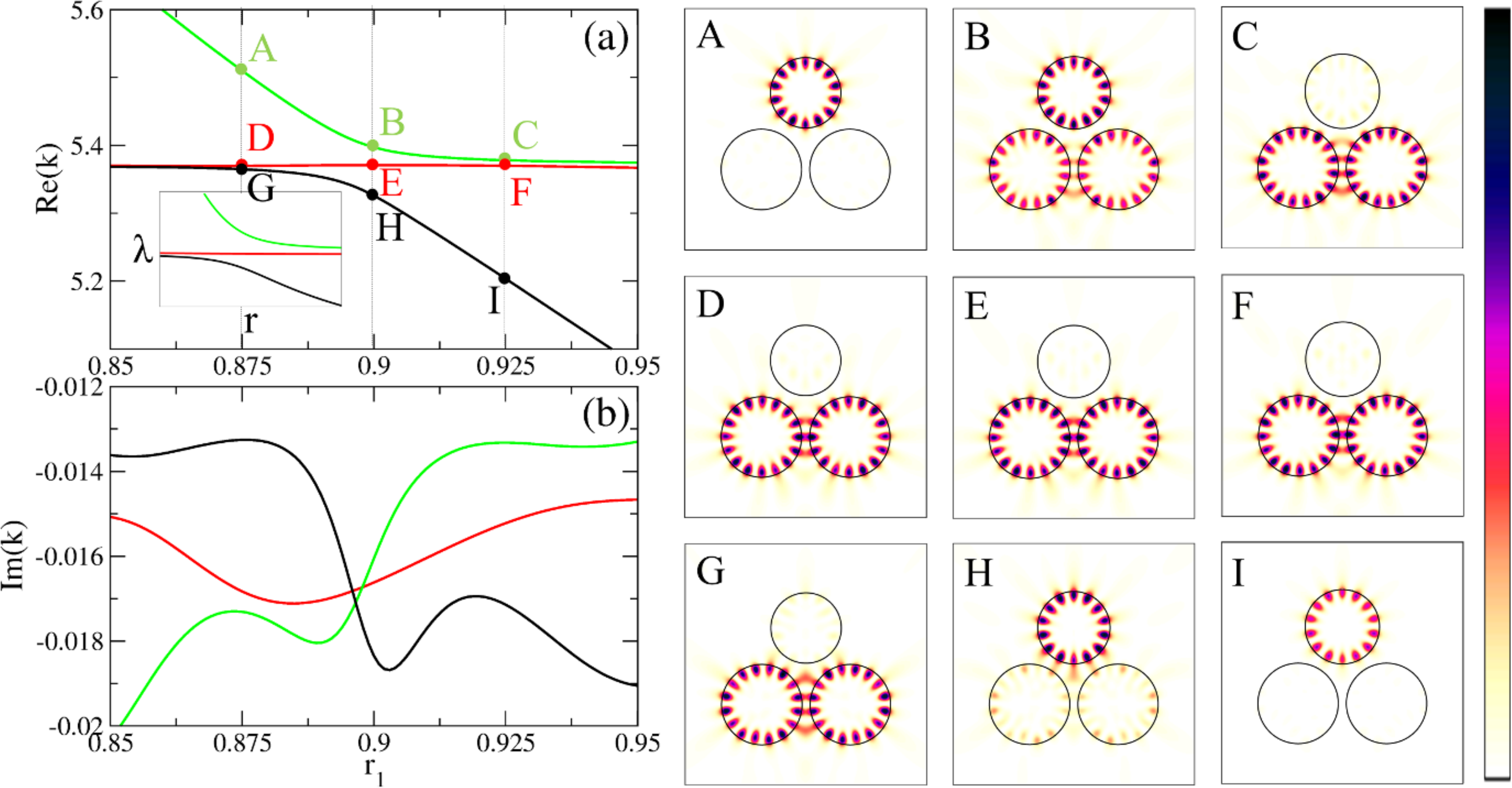}
\caption{(color online). (a) The real and (b) imaginary parts of complex $k$ of the resonant modes in the vicinity of RAC. The vertical dashed lines represent $r_1 = 0.875$, $0.9$, and $0.925$, respectively. The inset shows three eigenvalues $\lambda$ as a function of $r$ in Eq.~(3). The figures A-I represent the normalized intensity patterns corresponding to the resonant modes A-I.
}
\label{fig6}
\end{center}
\end{figure*}

\subsection{Matrix model}

As long as the coupling is not too strong, the eigenvalues and eigenstates of matrix models well explain the results of resonant modes in three disks of an isosceles triangular configuration as well as the equilateral triangular configuration. Without loss of generality, we only consider even-parity modes because the even and odd parity modes are independent. For the isosceles triangular case, there is no degeneracy between even and odd parity modes, unlike the equilateral triangular case. The local behaviors of WGM-type resonant modes of a single and coupled microdisks are also well explained by the simple $3 \times 3$ matrix $H_i$ of the form

\begin{equation}
H_{i} = \left(
\begin{array}{ccc}
\epsilon_1 & -c & -c \\
-c & \epsilon^{E}_{2} & 0 \\
-c & 0 & \epsilon^{E}_{3} \\
\end{array}
\right),
\label{matrix_2}
\end{equation}
where $\epsilon_{1}$ corresponds to a single WGM-type resonant mode on the top microdisk and $\epsilon^{E}_{2,3}$ do coupled WGM-type resonant modes with even parity about a vertical axis on the coupled microdisks, respectively, which can be obtained by $2 \times 2$ sub-matrix diagonalization for the coupled microdisks. When $c=0$, two corresponding eigenstates have even and odd parities about horizontal axis in coupled microdisks, respectively. For simplicity's sake, we assume all the coupling strengths between the resonant modes in a single microdisk and those in coupled microdisks are the same.

By setting $\epsilon_{1}=1/r$, we can obtain eigenvalues $\lambda$ of the matrix as a function of $r$ when $c=0.1$ and $\epsilon^{E}_{2,3}=1.0 \pm 0.01$ for the nearly degenerated coupled WGMs. As shown in the inset of Fig.~\ref{fig6} (a), the eigenvalues as a function of $r$ behaves similar to numerical results in Fig.~\ref{fig6} (a). It is noted that the complex resonant modes in microcavities are usually described by non-Hermitian matrices but Hermitian matrices such as Eq.(\ref{matrix_1}) and Eq.(\ref{matrix_2}) well describe our results because WGMs are low lossy resonant modes, i.e., the weak coupling case. It is also noticed that there are RACs with imaginary value crossings in Fig.~\ref{fig6}, however, RC with imaginary value avoided crossings could be observed as the loss of the system increases.

\section{Summary}

We have studied characteristic properties of whispering gallery modes in three microdisks of equilateral and isosceles triangular configurations. The coupled whispering gallery modes are explained by the symmetry of the systems, i.e., discrete rotational symmetry of the equilateral triangle and reflection symmetry of isosceles triangle. The corresponding simple matrix models which well explain the resonant modes behaviors in triple microdisks have been proposed. We expect the coupled whispering gallery modes in more complex multiple microdisks can be understood by the symmetric considerations and corresponding matrix models.

\section*{Funding}

J.-W.R was supported by Project Code (IBS-R024-D1).
S.R. was supported by the National Research Foundation of Korea (NRF) grant funded by the Korean government (MSIT) (No. 2017R1A2B4012045 and No. 2017R1A4A1015565).

\section*{Disclosures}

The authors declare no conflicts of interest.

\end{document}